\documentclass[preprint]{IEEEtran}

\usepackage{cite}
\usepackage{amsmath,amssymb,amsfonts}
\usepackage{algorithmic}
\usepackage{graphicx}
\usepackage{textcomp}
\usepackage{xcolor}
\def\BibTeX{{\rm B\kern-.05em{\sc i\kern-.025em b}\kern-.08em
    T\kern-.1667em\lower.7ex\hbox{E}\kern-.125emX}}

\usepackage{pgfplots}
\pgfplotsset{compat=1.18, width=10cm}
\usepackage{array} 

\begin{document}

\title{\textsc{InsureConnect:} Blockchain and Digital Identity for the Property Insurance Market}

\author{\IEEEauthorblockN{João Eduardo Travassos \ \ \ \ \  Miguel Correia}\\
\IEEEauthorblockA{\textit{INESC-ID, Instituto Superior Técnico, Universidade de Lisboa} -- Lisboa, Portugal} \\
\{joaotravassos,miguel.p.correia\}@tecnico.ulisboa.pt}

\maketitle

\begin{abstract}
This paper presents \textsc{InsureConnect}, a blockchain-based system for improving transparency, authentication, and auditability in property-insurance workflows after natural disasters. The system combines Self-Sovereign Identity (SSI), Decentralized Identifiers (DIDs), Verifiable Credentials (VCs), satellite imagery, Hyperledger Fabric, and IPFS to register identities, insurance contracts, and damage claims. Property images are stored off-chain in IPFS, while content hashes and signed records are maintained on a permissioned blockchain.
Users interact with the system through a desktop application, while chaincode enforces role-based access control and validates digital signatures. The prototype was evaluated under concurrent request loads from 50 to 3000 requests, measuring latency, throughput, and dropped connections. The results indicate that the system sustains increasing throughput under load, although latency rises and dropped connections appear at higher concurrency levels.
\end{abstract}

\section{Introduction}

Blockchain technology, first introduced in 2009 by \textit{Satoshi Nakamoto} in the Bitcoin whitepaper, has gained significant traction due to its diverse applications \cite{nakamoto2009bitcoin,peck2017blockchains}. As a decentralized peer-to-peer distributed ledger, blockchain eliminates the need for trusted third parties in online transactions. Over the years, its potential has expanded beyond cryptocurrency, leading to the development of private and permissioned blockchains that enable organizations to use this technology in contexts where trust is not inherent among parties.

This evolution also gave rise to \textit{Self-Sovereign Identity} (SSI), a transformative approach to identity management that empowers individuals to control their personal data without relying on centralized authorities \cite{hardman2022sovereignidentity}. Using Decentralized Identifiers (DIDs) and Verifiable Credentials (VCs), SSI leverages blockchain to provide a secure and privacy-preserving method for verifying identitie \cite{reed2020decentralized,w3c-vcdata-model,verifiable_credential_2023}. This innovation allows users to disclose only necessary information, fostering trust while maintaining privacy, making it particularly relevant in sectors like real estate and insurance.

Natural disasters, such as floods and earthquakes, pose significant challenges, often resulting in property damage and financial losses. Addressing these challenges requires innovative solutions that streamline the claim process and ensure trust between affected individuals and insurance companies. This paper presents \textsc{InsureConnect}, a system that registers entities, insurance contracts, and damage claims after natural disasters using Decentralized Identifier documents, satellite imagery, SSI, and blockchain technologies.

\textsc{InsureConnect} enhances the reliability and transparency of the insurance process through a blockchain-based framework. This shared ledger supports trust between clients seeking financial assistance and insurance companies processing claims. Although primarily aimed at natural disaster scenarios, \textsc{InsureConnect} has the potential to adapt to various markets, including auto insurance, serving as a versatile model for insurance-related transactions.

Built on a permissioned Hyperledger Fabric blockchain, \textsc{InsureConnect} uses DIDs and VCs to verify client identities and proof of property ownership. The InterPlanetary File System (IPFS) stores satellite images off-chain to avoid overloading the blockchain. The system provides three key services: DID Document registration, Insurance Contract registration, and Damage Claim registration. Each entity must first register with a DID Document, after which clients and insurance companies can establish insurance contracts based on property images. When a client experiences damage, they submit a claim, which the insurance company updates throughout the claim-handling process. Through this approach, \textsc{InsureConnect} aims to improve traceability, authentication, and auditability in property-insurance workflows after natural disasters.

\section{Background}

Blockchain\cite{dinh2018untangling} is a transparent, immutable, distributed, and decentralized ledger technology that securely links blocks of transactions using cryptographic techniques. It ensures data integrity and security, making it resistant to tampering and attacks. Blockchains can be public or private. Public blockchains, like Bitcoin\cite{nakamoto2009bitcoin} and Ethereum\cite{buterin2014ethereum}, are open to anyone and use consensus mechanisms such as proof-of-work to validate transactions, while private blockchains restrict access to authorized participants, providing greater control, faster transaction processing, and customizable permissions for different users.

Hyperledger Fabric\cite{hyperledger_fabric_2024} is a permissioned blockchain framework developed for enterprise solutions, featuring a modular architecture and plug-and-play consensus. It enables multiple channels, each representing a separate blockchain with its own ledger, allowing for private communication between specific members. Authentication and identity management are handled through cryptographic certificates issued by a Membership Service Provider (MSP). The transaction process in Fabric follows an execute-order-validate model, which includes endorsing transactions, ordering them into blocks, and validating them to ensure the integrity and consistency of the blockchain.

Digital identity involves verifying attributes beyond authentication, such as education or residency. Traditional systems require multiple siloed accounts, leading to privacy and security issues. \textit{Self-Sovereign Identity} (SSI) \cite{hardman2022sovereignidentity} uses blockchain to give individuals full control over their digital identity without relying on central authorities. With \textit{Decentralized Identifiers} (DIDs) \cite{reed2020decentralized} and \textit{Verifiable Credentials} (VCs) \cite{verifiable_credential_2023}, SSI enables privacy-preserving identity management, reducing intermediary dependence and minimizing data breaches. Decentralized ledgers support a user-centric trust model based on cryptographic proofs rather than centralized authorities.

The InterPlanetary File System (IPFS) \cite{benet2014ipfs} is a decentralized peer-to-peer file system designed for high-throughput content-addressed storage. It connects nodes using a Distributed Hash Table (DHT) for peer discovery and uses BitSwap, a protocol for exchanging data blocks between nodes, to incentivize content sharing. IPFS constructs a Merkle Directed Acyclic Graph (Merkle DAG) to link data cryptographically, ensuring content addressing, tamper resistance, and deduplication. This architecture ensures efficient file storage, content availability, and resilience without relying on a central authority or a single point of failure, making IPFS a robust solution for distributed storage.

\section{Related Work}

Blockchain has been increasingly studied in the insurance sector as a mechanism for improving transparency, traceability, automation, and auditability. A recent systematic literature review identifies blockchain as a promising technology for insurance workflows, while also noting that adoption remains limited by technical, organizational, regulatory, and economic challenges \cite{dominguez2024blockchainInsurance}. These challenges are particularly relevant for claims management, where several parties must agree on policy validity, evidence authenticity, claim status, and payment decisions.

Several works propose blockchain-based architectures for insurance claims. Chen et al.\ propose a car-insurance claims system based on blockchain, IPFS, Hyperledger Fabric, digital signatures, and smart contracts, aiming to reduce fraud, improve traceability, and lower the cost of storing claim-related data \cite{chen2023blockchainIPFSCarInsurance}. Their work is close to \textsc{InsureConnect} in its use of a permissioned blockchain and IPFS for claim evidence, but it focuses on car-insurance scenarios. \textsc{InsureConnect} instead targets property insurance after natural disasters and uses satellite imagery to document the state of insured property before and after a claim.

Self-Sovereign Identity (SSI) has also been explored in insurance-related blockchain systems. Farao et al.\ present INCHAIN, a cyber-insurance architecture that combines blockchain, smart contracts, and SSI to support traceability, process automation, and customer identification \cite{farao2024inchain}. This work demonstrates the relevance of SSI in insurance, especially for authentication and fraud reduction. However, INCHAIN focuses on cyber-insurance workflows, whereas \textsc{InsureConnect} applies DIDs and VCs to property-insurance contracts, proof of ownership, and post-disaster claims.

Natural-disaster insurance is also related to parametric insurance, where payouts are triggered by predefined indicators rather than traditional damage assessment. Steinmann et al.\ propose a framework for natural-hazard parametric insurance and discuss the importance of reducing basis risk, i.e., the mismatch between parameter-based payouts and actual damages \cite{steinmann2023naturalHazardParametric}. Rabehaja et al.\ address blockchain-based parametric insurance under multiple sources of truth, highlighting the risk of relying on a single data source for payout decisions \cite{rabehaja2024parametricInsurance}. Hao et al.\ further explore blockchain-enabled parametric insurance using remote sensing and IoT data, with privacy-preserving claim verification based on zero-knowledge proofs \cite{hao2025privacy}. These works are relevant because they show how external evidence sources can support insurance automation. In contrast, \textsc{InsureConnect} does not automate payouts through parametric triggers; instead, it records and authenticates contract and claim evidence so that clients, insurers, and auditors can verify the claim history.

The use of blockchain for remote-sensing image verification is another related area. Liu and Chang propose a blockchain-based method for retrieval and verification of remote-sensing images using Hyperledger Fabric and IPFS \cite{liu2024remoteSensingVerification}. Their work supports the idea that large geospatial images should be stored off-chain while hashes and metadata are maintained on-chain. \textsc{InsureConnect} follows a similar storage principle, but applies it to insurance evidence: satellite images are stored in IPFS, while blockchain records link those images to identities, contracts, claims, timestamps, and signatures.

Blockchain has also been studied in land-registry systems, which are relevant because property insurance depends on reliable ownership and property-state information. Several land-registry initiatives have explored blockchain in countries such as Sweden \cite{ICA-IT:BlockchainLandregistry}, India \cite{BlockchainGovIndia:LandRegistration}, Honduras \cite{BorgenProject:BlockchainLandRegistry}, Japan \cite{NikkeiAsia:JapanPropertyRecords}, and Brazil \cite{lemieux2018title}. Academic proposals in this area typically focus on registering property records, verifying ownership, and recording transfers through smart contracts \cite{khan2020blockchain, sahai2020smart, shuaib2020blockchain,henriques2025decentralised}. These systems improve the integrity of property records, but they generally do not address the full insurance lifecycle or integrate SSI-based identity verification with claim evidence.

Overall, existing work has explored blockchain in insurance, SSI for insurance identification, parametric disaster insurance, remote-sensing verification, and blockchain-based land registries. \textsc{InsureConnect} combines these directions in a single property-insurance workflow: participants are identified through DIDs, ownership can be supported through VCs, contracts and claims are recorded on a permissioned blockchain, and satellite imagery is stored off-chain through IPFS. The proposed system therefore focuses on authenticated and auditable evidence management for property-insurance claims after natural disasters.

\section{\textsc{InsureConnect} Design and Implementation}

\textsc{InsureConnect} is a system designed to help insurance companies and clients register, manage, and resolve insurance contracts and property damage claims, particularly for large terrains visible via satellite imagery. It focuses on natural disaster scenarios such as floods, earthquakes, and typhoons. Participants must register on the platform to establish contracts and submit claims. The authenticity of the data is ensured through digital signatures, allowing third-party verification. The media is stored in IPFS to achieve decentralization. 

\subsection{Participants and Roles}

Participants in the system are assigned roles that determine their interactions. The four main roles are:

\begin{itemize}
    \item \textit{Higher Authority} - A regulatory body responsible for creating and maintaining DIDs for insurance companies, ensuring the integrity and authenticity of their identities.
    \item \textit{Insurance Company} - Provides financial protection, manages contracts, and processes claims, especially during natural disasters.
    \item \textit{Client} - Holds a property insurance plan, submits damage claims, and updates contracts.
    \item \textit{Auditor} - Independently verifies financial records, with read-only access to disputed records for audit purposes.
\end{itemize}

Participants must be registered in two places: the blockchain ledger and Fabric's Membership Service Provider (MSP). Ledger registration creates a DID Document through chaincode and stores it permanently on the blockchain. MSP registration is separate from the ledger and is performed through the Fabric certificate authority, which issues the certificate and private key required to submit transactions. The Higher Authority is pre-assigned and registers Insurance Company users; Insurance Companies then register their Clients and Auditors.

\begin{figure}[t]
\centering
\includegraphics[width=1\linewidth]{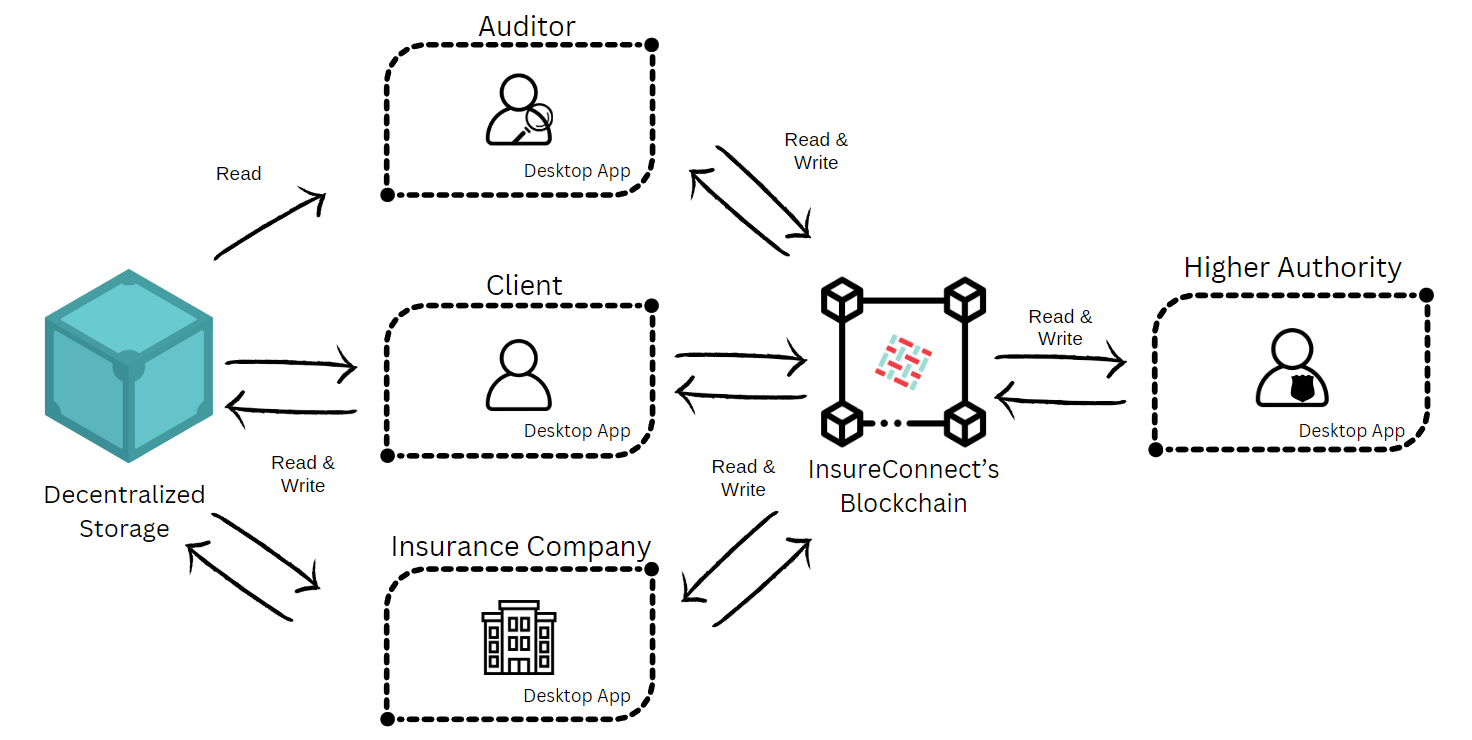}
\caption{InsureConnect roles and permissions} 
\label{fig:InsureConnectRolesPermissions}
\end{figure}

Each role has specific permissions (see Figure \ref{fig:InsureConnectRolesPermissions}):

\begin{itemize}
    \item \textit{Insurance Companies}: Can create, update, and read contracts and claims.
    \item \textit{Client}: Can update contracts and manage their claims.
    \item \textit{Auditor}: Have read-only access to disputed records for verification.
\end{itemize}

These roles ensure secure and controlled management of insurance operations.

\subsection{Architecture}

\textsc{InsureConnect} uses a private, permissioned Hyperledger Fabric blockchain to store records, a decentralized storage network for images, and a desktop application through which participants interact with the system.

\textsc{InsureConnect} is designed as a decentralized, containerized system using Docker for consistent deployment and orchestration. All components, blockchain nodes and IPFS storage, are packaged in Docker containers, ensuring portability, scalability, and service isolation. Docker simplifies network management, allowing for seamless node addition or removal. The architecture has three layers: the blockchain network, decentralized storage, and client interface. Each component and their interaction is detailed below.

\begin{itemize} 
    \item \textit{\textsc{InsureConnect} Blockchain} - Hyperledger Fabric stores DID Documents, Insurance Contracts, and Claims. This permissioned blockchain ensures that only relevant users can access their data, providing privacy and security. Chaincode enforces access controls and smart contract execution, ensuring trust, transparency, and data immutability.

    \item \textit{Decentralized Storage} - A peer-to-peer network, using IPFS, stores satellite images of the property, reducing blockchain storage costs. The blockchain holds content hashes linking to IPFS, ensuring tamper-resistance and easy access to off-chain data.

    \item \textit{Desktop App} - The app allows users to interact with the blockchain and decentralized storage. Users can securely sign transactions, submit queries, and manage files in a seamless and secure interface.
\end{itemize}

\begin{figure}[t] 
    \centering 
    \includegraphics[width=1\linewidth]{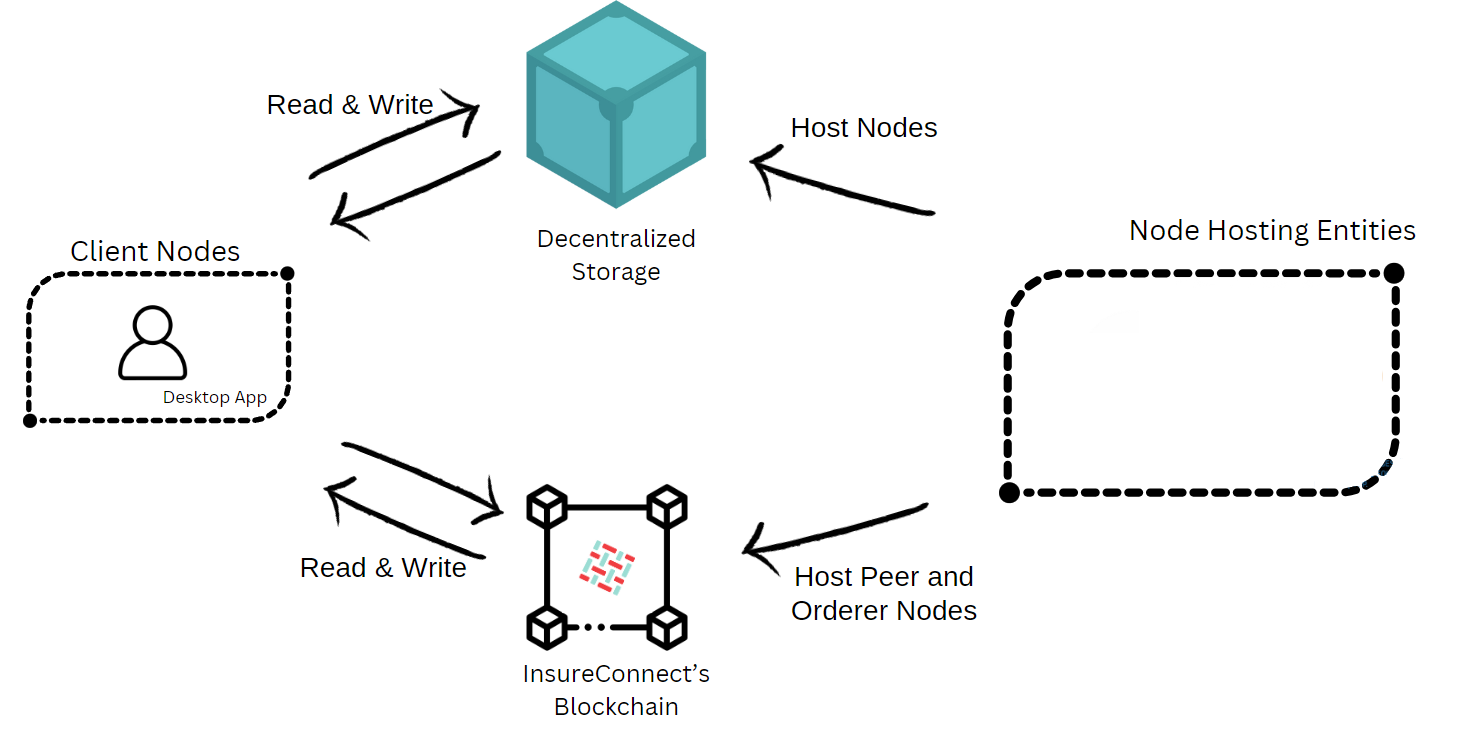} \caption{\textsc{InsureConnect} architecture} 
    \label{fig:InsureConnectArchitecture} 
\end{figure}

Figure \ref{fig:InsureConnectArchitecture} shows the architecture of \textsc{InsureConnect}.

The solution decentralizes record storage and verification among authorized participants in a permissioned network. All records, such as DID documents, insurance contracts, and damage claims, are stored on the blockchain. Due to the high cost of on-chain storage, the system uses IPFS, a decentralized, tamper-resistant, peer-to-peer storage network. IPFS nodes, hosted alongside Hyperledger Fabric nodes, allow participants to store and access data via the desktop app.

\subsection{Blockchain}

The blockchain chosen for this project is Hyperledger Fabric \cite{Dhillon2017} from the Linux Foundation. It was selected because it meets the project's requirements, such as identity management.

The blockchain should be hosted by at least five different entities to ensure fault tolerance. Each entity will host a peer node, and 3 of them will also host an orderer node, resulting in a total of 8 nodes. External applications connecting to peer nodes to submit transactions will act as client nodes. This setup ensures network resilience, as peer nodes maintain the ledger and endorse transactions, even if some nodes go offline. The distributed structure supports scalability, allowing the network to handle more transactions as it grows.

The orderer nodes use the \textit{Raft} consensus mechanism \cite{RAFTongaro2014search}, chosen over Kafka and Solo due to its simplicity and ease of maintenance. Raft handles leader election and fault tolerance effectively, making it the preferred choice for Fabric, while Kafka adds unnecessary complexity with its need for a ZooKeeper cluster. Solo, designed for testing only, lacks fault tolerance and is not suitable for production.

Kafka provides strong consistency and fault tolerance, but its reliance on ZooKeeper adds complexity and can affect the availability of the system during failures or maintenance. According to the CAP Theorem \cite{CAPTHEOREM}, distributed systems can provide only two of three properties: consistency, availability, and partition tolerance. Raft is a CP protocol, ensuring consistency during network partitions but sacrificing availability during leader reelections. Kafka also follows the CP model, but ZooKeeper's setup increases operational complexity, impacting even more the availability.

There are three types of nodes: peer, orderer, and client nodes. Client nodes are used by participants to submit transactions and queries to the network. Each participant uses two distinct credential sets. The first is the DID key pair, used to sign the business content submitted to the ledger, such as DID Documents, contracts, and claims. The second is the Fabric MSP credential, composed of a certificate and private key, used by the client application to authenticate the transaction or query issuer to Hyperledger Fabric. This distinction is especially important for read operations, where no signed business object is submitted but the chaincode must still verify which participant issued the query.

In addition to peer and orderer nodes, at least one entity must host a certificate authority (CA) to register participants and issue transaction signing credentials.

\subsection{Ledger}

Each peer node will have one chaincode deployed, to which the client nodes will submit transactions. All objects in the network are stored in a single ledger.

\begin{figure}[t] \centering \includegraphics[width=1\linewidth]{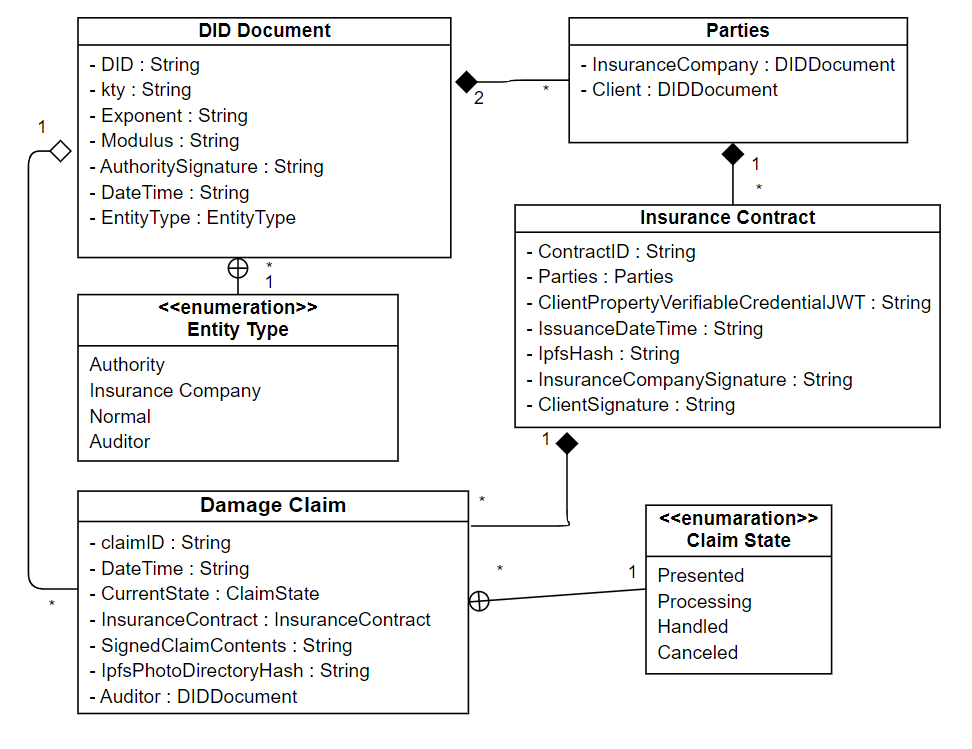} \caption{\textsc{InsureConnect} UML class diagram.} \label{fig:InsureConnectUMLClassDiagram}
\end{figure}

The chaincode represents three main objects (see Figure \ref{fig:InsureConnectUMLClassDiagram}): DID documents, insurance contracts, and damage claims. These objects, which contain crucial data, must be accompanied by a signature in JSON format when submitted. The main objects are:

\begin{itemize} 
\item \textit{DID Documents} - Stored in the ledger to prevent divergent results from external data updates, ensuring consistent blockchain progress. These documents are vital for identifying and verifying who performs write operations. They include the entity's identity (`did'), an RSA public key (composed of `exponent' and `modulus'), and additional fields like `kty' (key type), `DateTime' (creation time), and `EntityType' (defining role and access permissions). Peers verify the signature before endorsing transactions.

\item \textit{Insurance Contracts} - Represent agreements between two entities. Each contract contains DIDs for both parties, an IPFS link to property images, verifiable credentials, creation timestamp, and both parties' signatures. The contract is valid only after both parties sign. Any updates erase the signatures, which require the re-signing for validity.

\item \textit{Damage Claims} - Allow clients to seek compensation for losses. Each claim includes a reference to the client, insurance contract, an IPFS link, creation timestamp, claim state (starting with `PRESENTED'), and the signature of the issuer (client or insurance company). Claims progress through predefined states: `PRESENTED', `PROCESSING', `HANDLED', and `CANCELED', without reverting to previous stages. Each update requires a new signature and updated values. 
\end{itemize}

Figure \ref{fig:InsureConnectUMLClassDiagram} shows the UML class diagram of these ledger objects.
Note that while the `kty' field exists in DID documents, it is unused in this prototype, since only RSA signatures are accepted. RSA was chosen due to system constraints. Hyperledger Fabric runs on Java 11, which limited the use of faster algorithms such as EdDSA, which is available only in Java 16 or higher. As a result, RSA was the first algorithm successfully implemented.

\subsection{Decentralized Storage}

As mentioned above, the Interplanetary File System (IPFS) is used to store all content related to property damage, specifically images taken before and after a claim is submitted to document damage. When creating an insurance contract or claim object, an IPFS link is required pointing to an IPFS directory containing these photos.
IPFS plays an important role due to the large storage space required for images, which cannot be stored directly on Fabric's blockchain. To maintain a decentralized solution, IPFS was chosen as the off-chain decentralized storage system for \textsc{InsureConnect}.

Other options, such as BitTorrent \cite{cohen2003incentivesBITTORRENT} and Swarm \cite{swarm_whitepaper}, were considered. Swarm was discarded because it is tightly integrated with the Ethereum network. BitTorrent, while competitive in performance and efficiency, loses in decentralization and content immutability. BitTorrent relies on centralized trackers to assist in peer discovery, whereas IPFS uses a fully decentralized distributed hash table (DHT). Furthermore, IPFS's content-addressing ensures data immutability, as altering a file would change its hash, making it useful for proving that images have not been tampered with.

The prototype uses the public IPFS network, which means that uploaded images may be retrievable by any party that obtains the corresponding content identifier. Therefore, the IPFS link should not be treated as an access-control mechanism. In a production deployment, sensitive images should be encrypted before being added to IPFS, with decryption keys shared only with authorized parties such as the client, the insurance company, and assigned auditors.

A key issue with IPFS is its garbage collector, which removes unused data. To prevent this, content can be pinned, ensuring that it is not deleted.

An alternative approach could involve creating a private national IPFS network to ensure privacy. However, this would require additional entities and storage units. Relying on the five current entities hosting the Fabric blockchain would result in a centralized system, which undermines true decentralization, as the failure of one or two nodes could compromise the network.

\subsection{Desktop App}

As mentioned above, the desktop app will be used by all users to interact with the network. The app does not differentiate between user types like ``Insurance Company" or ``Client." Instead, the chaincode on the blockchain handles role-based access control. Thus, the app provides every user with all available interaction options. The user identities and signatures submitted to the blockchain are used to verify transactions.

\textsc{InsureConnect}'s desktop app is divided into four major components based on functionality:

\begin{itemize} 
\item \textit{Identity Management Menu} – Responsible for creating DID documents, retrieving identity data, and updating verifiable credentials. 
\item \textit{Insurance Contract Management Menu} – Handles the creation of insurance contracts, updating client signatures, modifying contracts, and retrieving contract data. 
\item \textit{Claim Management Menu} – Enables creating, updating, and retrieving claim data. \item \textit{Add Public and Private KeyPairs Menu} – Allows users to upload their public and private keypairs, issued by Fabric's MSP, so the app can sign transactions for recognition by the blockchain. 
\end{itemize}

The ``Add Certificate and Private Key" menu is essential for entities to upload their certificate/private key pair, enabling the app to sign transactions and allowing the blockchain to recognize and verify them.

\subsubsection{WebApp}

Although the desktop application was ultimately chosen as the final solution, the original approach involved a web application hosted by each node. However, this solution was later discarded due to the Read operation issue described earlier. In the web app solution, users would submit the transaction content and its corresponding signature through the web interface. The web app's host would then build the transaction request and submit it using admin privileges. While Write operations would remain unchanged, Read operations would become indistinguishable, making it impossible to identify who submitted them. This issue arose because all transaction requests originated from the same user, the node that received the transaction contents via the web app and built the transaction request to send to Fabric. Due to privacy concerns, the web app solution was abandoned in favor of a more privacy-respecting desktop application.

\section{\textsc{InsureConnect} Operations}
\label{section:insureConnect Operations}

\begin{figure*}[t]
\centering
\includegraphics[width=0.75\linewidth]{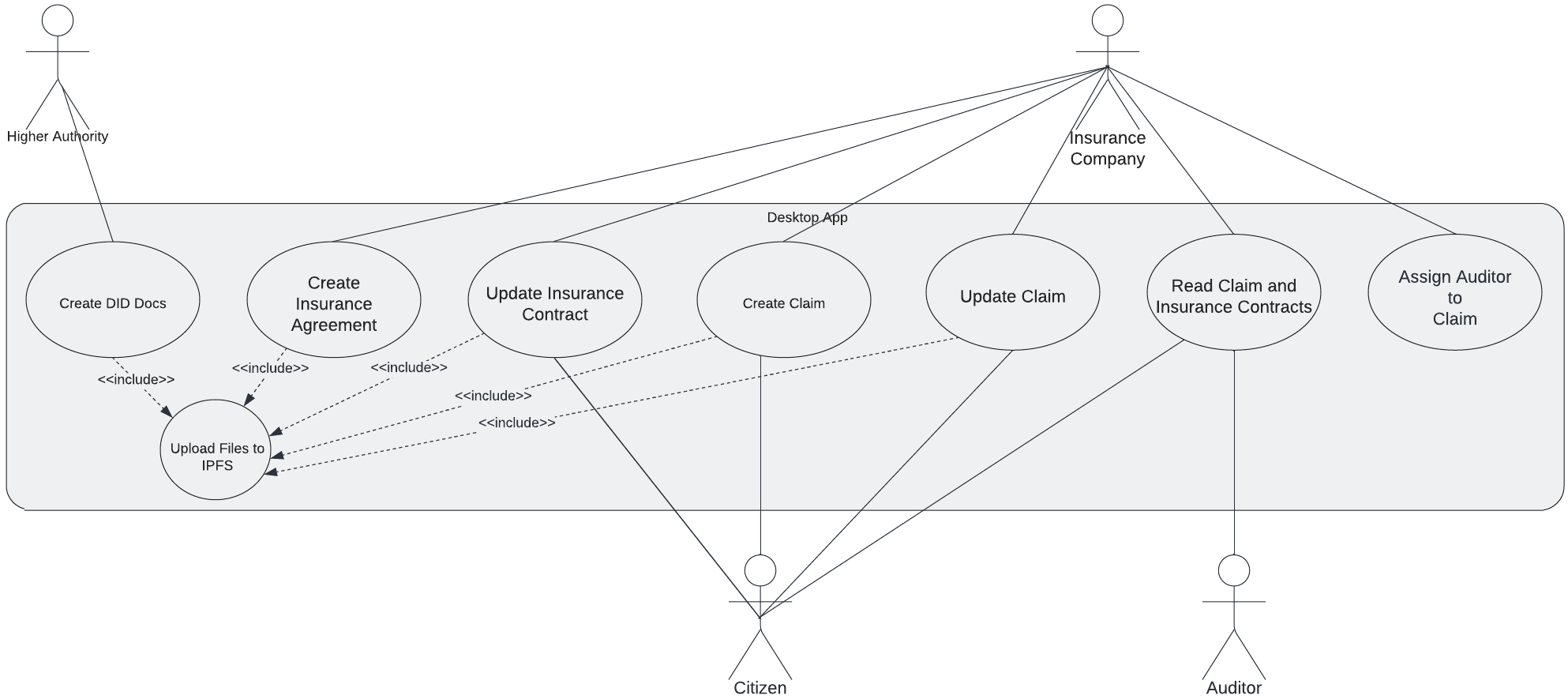}
\caption{\textsc{InsureConnect}'s operations diagram} 
\label{fig:InsureConnectInDepthArchitecture}
\end{figure*}

This section describes the supported operations in the chaincode. It is divided into two subsections where each one addresses different objects. The first subsection talks about identity object methods, specifically focusing on DID documents. The second subsection details methods for interacting with insurance contracts and managing claims. Figure \ref{fig:InsureConnectInDepthArchitecture} shows all possible operations for every role.
This section also aims to guide the reader through the entire process, from the creation of a DID Document to the resolution of a Claim.

\subsection{Identity Management}

Within this menu, there are two further options: Create a DID Document and Check a DID Document.

To create a DID Document, the following fields must be provided:
\begin{itemize}
    \item \textit{Decentralized Identifier (DID)} - This field holds the identity of the entity on the platform.
    \item \textit{Exponent} - This field holds the exponent of an RSA key.
    \item \textit{Modulus} - This field contains the modulus of an RSA key.
    \item \textit{Date Time} - This field holds the date and time at which the transaction was signed and submitted to the network.
    \item \textit{Entity Type} - This field holds the type of user this entity will be.
    \item \textit{Authority Signature} - This field holds the signature of the user who issued the transaction and submitted the DID Document creation request.
\end{itemize}

Once the transaction is submitted to the Fabric network, the signature is verified. A Higher Authority user signature is required to create Insurance Company users and an Insurance Company user signature is needed to create Client and Auditor users. If the signature is valid, a new DID Document is created and stored in the ledger.

\begin{figure}[t]
\centering
\includegraphics[width=1\linewidth]{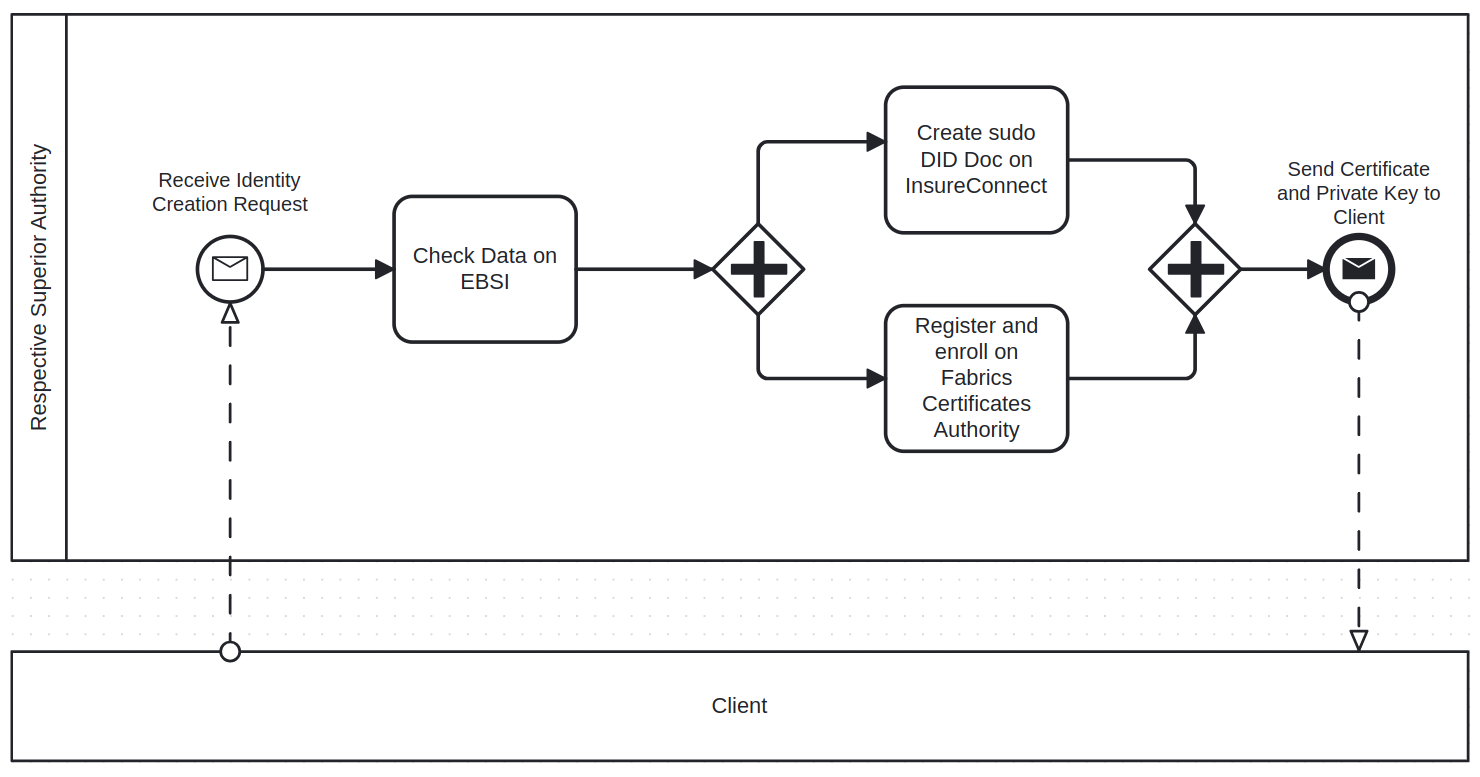}
\caption{BPMN diagram of the identity creation process} 
\label{fig:createIdentitybpmn}
\end{figure}

At this stage, an additional procedure is needed: the registration and enrollment of the user in Fabric's MSP. As mentioned above, this is done by users with admin permissions. These admin permissions are granted only to the Higher Authority and Insurance Companies. Once registration and enrollment are complete, the generated public/private key pair is sent back to the newly created user, allowing them to load the credentials into their desktop app. Figure \ref{fig:createIdentitybpmn} illustrates the identity creation process.

To check a DID Document, the only required input is the DID that the user wishes to verify. The corresponding data is then retrieved and displayed.

\subsection{Insurance Contract Management}

Within this menu, there are four further options, Create an Insurance Contract,  Update an Insurance Contract,  Update the Client Signature and Check my Insurance Contracts.

To create an Insurance Contract, the following fields must be provided:
\begin{itemize}
    \item \textit{Insurance Company DID} - This field holds the identity of the insurance company.
    \item \textit{Client DID} - This field holds the identity of the client.
    \item \textit{Client Property Land Registration VC} - This field holds the client's VC issued by Land Registration Authority stating that the client is the owner of the property the contract will cover.
    \item \textit{IPFS Hash} - This field holds the link to an IPFS directory with the images of the property at the time of contract issuance.
    \item \textit{Date Time} - This field holds the date and time at which the transaction was signed and submitted to the network.
    \item \textit{Insurance Company Signature} - This field holds the signature of the insurance company issuing the transaction and submitting the Insurance Contract creation request.
\end{itemize}

Once the contract is created, the Client user must use the Update the Client Signature option to add their signature to the insurance contract, making it valid.

\begin{figure}[t]
\centering
\includegraphics[width=1\linewidth]{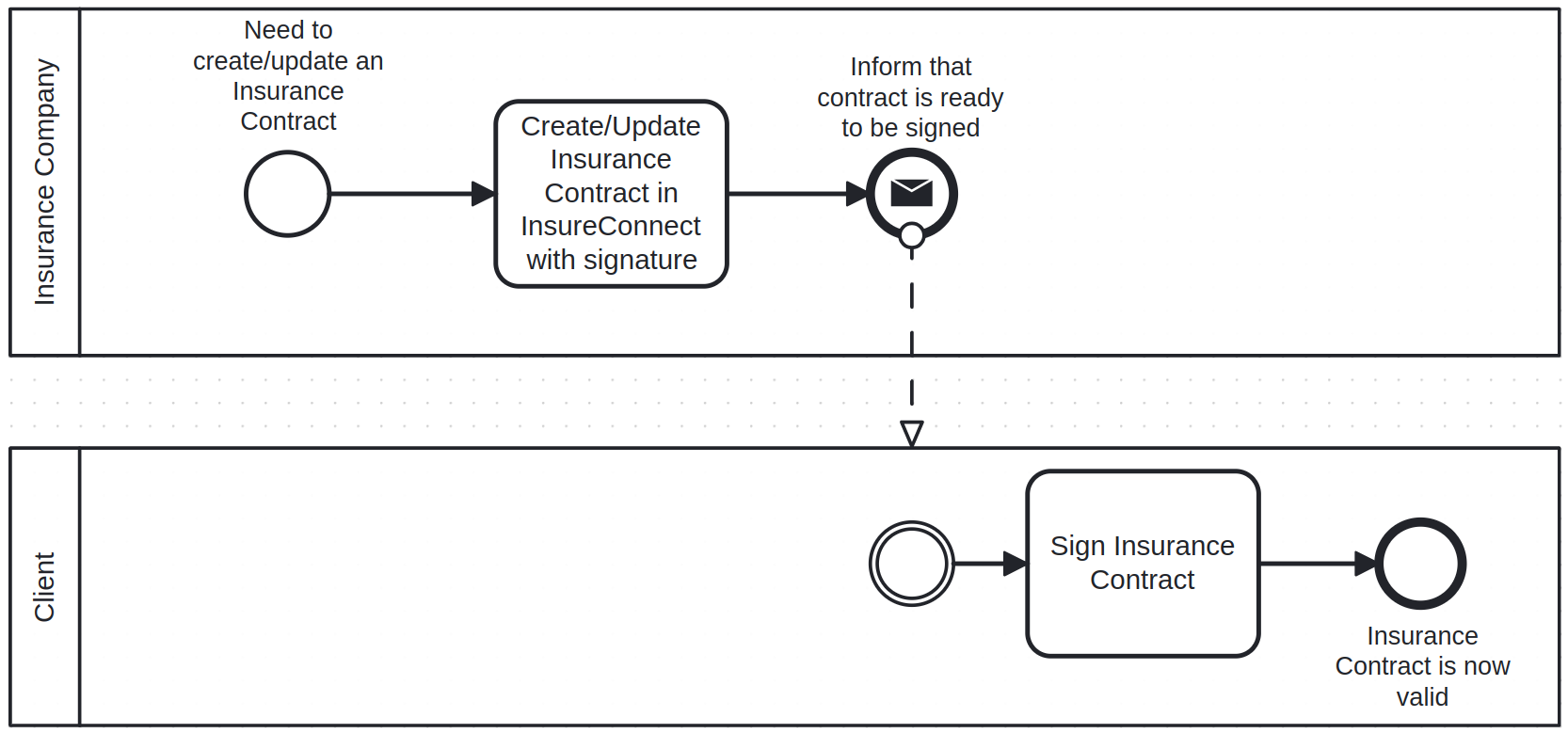}
\caption{BPMN diagram of the insurance contract creation/update} 
\label{fig:createUpdateInsuranceContractbpmn}
\end{figure}

An insurance contract can be updated, but only by the insurance company. The only field which is allowed to be updated is the IPFS Hash. Once it is updated, the Date Time field also is updated to reflect the time of the modification. The insurance company must submit a new signature along with the updated fields, and the client's signature is deleted. While the client does not update their signature, the contract is considered invalid. Once the client signature is added to the contract, the contract becomes valid again. Figure \ref{fig:createUpdateInsuranceContractbpmn} demonstrates the Insurance Contract creation/update process which happens to be the same.

To check an Insurance Contract, the user provides the DID whose contracts should be retrieved. Because read operations do not include a signed business object, the system uses the Fabric MSP identity to authenticate the query issuer. The desktop application signs the Fabric query with the user's MSP private key, and the chaincode obtains the issuer identity from the transaction context. If the issuer is authorized to access contracts associated with the target DID, the corresponding contracts are returned; otherwise, the request is rejected.

\subsection{Claims Management}

Within this menu there are three further options: Create a Claim,  Update a Claim, Assign an Auditor, and Check Claims.

To create a Claim, the following fields must be submitted:
\begin{itemize}
    \item \textit{Claim ID} - This field holds ID of the claim.
    \item \textit{Insurance Contract ID} - This field holds the ID of the insurance contract the claim will be covered by.
    \item \textit{IPFS Hash} - This fields holds the link to an IPFS directory with the images of the property at the time of claim issuance.
    \item \textit{Date Time} - This field holds the date and time at which the transaction was signed and submitted to the network.
    \item \textit{Contents Signature} - This field holds the signature of either the insurance company or the client, depending on who issued the transaction and submitted the Claim creation request.
\end{itemize}

The claim can be created by either the client or the insurance company.
The fields Claim State and Auditor are not filled during the creation, since they will be automatically filled in the object by the chaincode. By default, the claim's initial state will be PRESENTED, and the Auditor will be set to null. The starting stage must also be included in the signature.

\begin{figure}[t]
\centering
\includegraphics[width=1\linewidth]{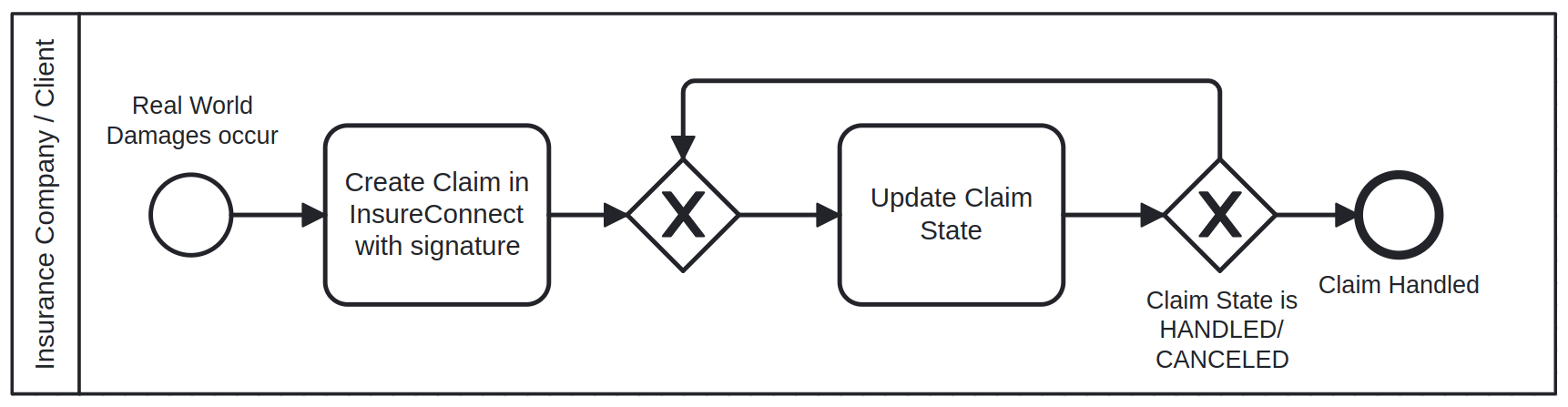}
\caption{BPMN diagram of the lifecycle of a claim} 
\label{fig:claimLifecycle}
\end{figure}

After the claim is created, both the client and the insurance company can use the Update a Claim option to update the claim. In addition, the insurance company can assign an auditor to the claim. The only editable fields are the Claim State and Auditor fields, respectively, for each function. The new updated claim state must progress for the claim to be handled and finished. Again, when either claim state or the auditor are updated, a new signature must be submitted by the one who updated it, along with the new updated contents. Figure \ref{fig:claimLifecycle} demonstrates the lifecycle of a claim.

To check a Claim, the user provides the target DID or claim identifier. As with contract queries, the desktop application signs the query using the user's Fabric MSP credentials. The chaincode then verifies whether the requester is authorized to access the claim. Clients and insurance companies can access claims associated with their contracts, while auditors can access claims assigned to them. If the requester is not authorized, the query is rejected.

\subsection{Solution Recap}

This section presented the \textsc{InsureConnect} system, whose objective is to establish a well-documented insurance process from the beginning to the end. From the moment the entities are created on the platform to the moment that a claim has finished its lifecycle. 

The section begins by establishing the roles and participants, Client, Insurance Company, Auditor, and Higher Authority, while explaining their access permissions. Following that, the architecture is laid out, detailing the active components involved: Hyperledger Fabric, IPFS, the desktop app, and the possible nodes for hosting the network. Then each component is explained in detail to justify every decision that affected the final product. Such decisions were, for example, the chosen consensus mechanism, Raft, the need for two different public/private keypairs, why are RSA signatures utilized instead of others, and why was a desktop app developed rather than an online web interface. 

To conclude this section, all of \textsc{InsureConnect}'s operations are described to provide insights into the features offered by the desktop app. The different lifecycles of the objects are also presented.

\section{Evaluation}


The evaluation focused on measuring how the system performs under different loads of concurrent requests. The performance of five functions was tested, each subject to request loads ranging from 50 to 3000 concurrent requests.

The evaluation measures the performance of five write operations: Create DID Document, Create Insurance Contract, Update Insurance Contract Signature, Create Claim, and Update Claim. These operations were selected because they exercise the most resource-intensive parts of the system, including signature verification, endorsement, ordering, validation, ledger updates, and the read operations required for data lookup. The results provide an initial basis for understanding the system's behavior under dynamic request loads.

Across the experiments, latency increased with the number of concurrent requests. Throughput also increased initially, but its growth slowed at higher request loads as system resources became saturated. At the highest loads, some requests failed due to dropped connections. Table \ref{tab:ErrorRateTable} reports the percentage of dropped connections for each operation.

The experiments were limited to 3000 concurrent requests because higher loads made the host machine unstable, producing invalid measurements or causing all connections to fail.

\begin{figure}[t]
    \centering
    \begin{minipage}{0.45\linewidth}
        \centering
        \begin{tikzpicture}[scale=0.40]
            \begin{axis}[title={Create DID Document Latency},
                xlabel={Number of Requests},
                ylabel={Latency (s)}]
                \addplot[
                color=blue,
                mark=square,
                ]
                coordinates {
                (50, 1.16) (100, 1.81) (250, 3.52) (500, 7.39) (1000, 11.86) (1500, 13.26) (2000, 18.73) (2500, 20.55) (3000, 24.54) 
                };
            \end{axis}
        \end{tikzpicture}
    \end{minipage}%
    \hspace{0.05\linewidth} 
    \begin{minipage}{0.45\linewidth}
        \centering
        \begin{tikzpicture}[scale=0.40]
            \begin{axis}[title={Create DID Document Throughput},
                xlabel={Number of Requests},
                ylabel={Throughput (Tx/s)}]
                \addplot[
                color=blue,
                mark=square,
                ]
                coordinates {
                (50, 37.85) (100, 44.80) (250, 50.44) (500, 52.04) (1000, 64.86) (1500, 77.06) (2000, 79.76) (2500, 80.89) (3000, 84.73) 
                };
            \end{axis}
        \end{tikzpicture}
    \end{minipage}
    \caption{Latency \& throughput of the create DID documents function}
    \label{fig:ltc_tgp_createDIDDoc}
\end{figure}
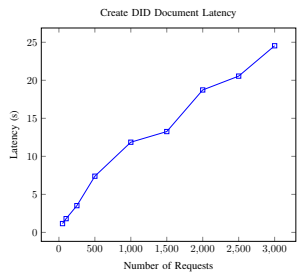

Figure \ref{fig:ltc_tgp_createDIDDoc} shows the latency and throughput of the Create DID Document operation. Latency increases almost linearly with the number of concurrent requests, from 1.16 s at 50 requests to 24.54 s at 3000 requests. Throughput also increases, but the growth rate slows at higher loads, indicating that the system approaches resource saturation.

Figure \ref{fig:ltc_tgp_createInsuranceContract} presents the results for the Create Insurance Contract operation. Latency increases with the request load, reaching 25.63s at 3000 concurrent requests. Throughput increases across the tested range, reaching 74.84 Tx/s at 3000 requests. However, this operation begins to show dropped connections at 2500 requests, as reported in Table \ref{tab:ErrorRateTable}, indicating that the system is approaching its resource limits.

\begin{figure}[t]
    \centering
    \begin{minipage}{0.45\linewidth}
        \centering
        \begin{tikzpicture}[scale=0.4]
            \begin{axis}[title={Create Insurance Contract Latency},
                xlabel={Number of Requests},
                ylabel={Latency (s)}]
                \addplot[
                color=blue,
                mark=square,
                ]
                coordinates {
                (50,1.77)(100,2.82)(250,4.80)(500, 7.77)(1000, 12.24) (1500, 15.57) (2000, 18.61) (2500, 21.02) (3000, 25.63)
                };
            \end{axis}
        \end{tikzpicture}
    \end{minipage}%
    \hspace{0.05\linewidth} 
    \begin{minipage}{0.45\linewidth}
        \centering
        \begin{tikzpicture}[scale=0.4]
            \begin{axis}[title={Create Insurance Contract Throughput},
                xlabel={Number of Requests},
                ylabel={Throughput (Tx/s)}]
                \addplot[
                color=blue,
                mark=square,
                ]
                coordinates {
                (50,25.37)(100,29.06)(250,34.49)(500,42.53)(1000, 54.23) (1500, 59.95) (2000, 65.43) (2500, 67.42) (3000, 74.84)
                };
            \end{axis}
        \end{tikzpicture}
    \end{minipage}
    \caption{Latency \& throughput of the create insurance contract function}
    \label{fig:ltc_tgp_createInsuranceContract}
\end{figure}

Figure \ref{fig:ltc_tgp_updateInsuranceContract} presents the results for the Update Insurance Contract Signature operation. Latency increases from 1.22s at 50 requests to 17.04 s at 2500 requests, followed by a slight decrease to 15.44s at 3000 requests. Throughput generally increases with load, although the measurements fluctuate at higher concurrency levels. These fluctuations suggest that the system is operating near saturation.

\begin{figure}[t]
    \centering
    \begin{minipage}{0.45\linewidth}
        \centering
        \begin{tikzpicture}[scale=0.4]
            \begin{axis}[title={Update Insurance Contract Signature Latency},
                xlabel={Number of Requests},
                ylabel={Latency (s)}]
                \addplot[
                color=blue,
                mark=square,
                ]
                coordinates {
                (50,1.22)(100,2.27)(250,3.46)(500, 4.18)(1000, 6.75) (1500, 8.82) (2000, 15.28) (2500, 17.04) (3000, 15.44)
                };
            \end{axis}
        \end{tikzpicture}
    \end{minipage}%
    \hspace{0.05\linewidth} 
    \begin{minipage}{0.45\linewidth}
        \centering
        \begin{tikzpicture}[scale=0.4]
            \begin{axis}[title={Update Insurance Contract Signature Throughput},
                xlabel={Number of Requests},
                ylabel={Throughput (Tx/s)}]
                \addplot[
                color=blue,
                mark=square,
                ]
                coordinates {
                (50,35.35)(100,31.04)(250,47.32)(500,65.86)(1000, 79.30) (1500, 85.6) (2000, 78.28) (2500, 85.62) (3000, 105.82) 
                };
            \end{axis}
        \end{tikzpicture}
    \end{minipage}
    \caption{Latency \& throughput of the update insurance contract signature function}
    \label{fig:ltc_tgp_updateInsuranceContract}
\end{figure}

Figure \ref{fig:ltc_tgp_createClaim} shows the results for the Create Claim operation. Latency increases from 1.83 s at 50 requests to 24.50 s at 3000 requests. Throughput increases up to 2500 requests, reaching 86.38 Tx/s, before decreasing to 77.02 Tx/s at 3000 requests. This decrease, together with the dropped connections reported in Table \ref{tab:ErrorRateTable}, indicates that this operation is affected by resource saturation at high concurrency levels.

\begin{figure}[t]
    \centering
    \begin{minipage}{0.45\linewidth}
        \centering
        \begin{tikzpicture}[scale=0.4]
            \begin{axis}[title={Create Claim Latency},
                xlabel={Number of Requests},
                ylabel={Latency (s)}]
                \addplot[
                color=blue,
                mark=square,
                ]
                coordinates {
                (50,1.83)(100, 2.32)(250,5.06)(500, 6.34)(1000, 7.95) (1500, 9.99) (2000, 15.14) (2500, 16.48) (3000, 24.50)
                };
            \end{axis}
        \end{tikzpicture}
    \end{minipage}%
    \hspace{0.05\linewidth} 
    \begin{minipage}{0.45\linewidth}
        \centering
        \begin{tikzpicture}[scale=0.4]
            \begin{axis}[title={Create Claim Throughput},
                xlabel={Number of Requests},
                ylabel={Throughput (Tx/s)}]
                \addplot[
                color=blue,
                mark=square,
                ]
                coordinates {
                (50,22.71)(100,34.96)(250,37.41)(500,50.78)(1000, 73.08) (1500, 78.62) (2000, 76.27) (2500, 86.38) (3000, 77.02)
                };
            \end{axis}
        \end{tikzpicture}
    \end{minipage}
    \caption{Latency \& throughput of the create claim function}
    \label{fig:ltc_tgp_createClaim}
\end{figure}

\begin{figure}[t]
    \centering
    \begin{minipage}{0.45\linewidth}
        \centering
        \begin{tikzpicture}[scale=0.4]
            \begin{axis}[title={Update Claim Latency},
                xlabel={Number of Requests},
                ylabel={Latency (s)}]
                \addplot[
                color=blue,
                mark=square,
                ]
                coordinates {
                (50, 1.40) (100, 2.08) (250, 3.99) (500, 4.16) (1000, 7.34) (1500, 7.71) (2000, 9.52) (2500, 13.56) (3000, 14.21) 
                };
            \end{axis}
        \end{tikzpicture}
    \end{minipage}%
    \hspace{0.05\linewidth} 
    \begin{minipage}{0.45\linewidth}
        \centering
        \begin{tikzpicture}[scale=0.4]
            \begin{axis}[title={Update Claim Throughput},
                xlabel={Number of Requests},
                ylabel={Throughput (Tx/s)}]
                \addplot[
                color=blue,
                mark=square,
                ]
                coordinates {
                (50, 30.06) (100, 38.40) (250, 47.78) (500, 70.64) (1000, 83.09) (1500, 103.7) (2000, 108.50) (2500, 109.69) (3000, 138.86) 
                };
            \end{axis}
        \end{tikzpicture}
    \end{minipage}
    \caption{Latency \& throughput of the update claim function}
    \label{fig:ltc_tgp_UpdateClaim}
\end{figure}
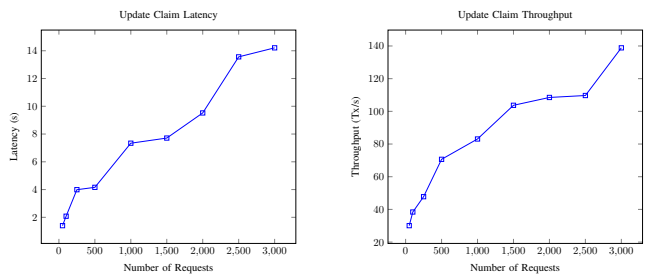

Figure \ref{fig:ltc_tgp_UpdateClaim} presents the results for the Update Claim operation. Latency increases from 1.40s at 50 requests to 14.21s at 3000 requests. Throughput also increases across the tested range, reaching 138.86 Tx/s at 3000 requests. Nevertheless, this operation has the highest dropped-connection rate at high load, reaching 20\% at 2500 requests and 30\% at 3000 requests.

\begin{table*}[t]
\centering
\caption{Error rate versus number of requests (in percentage)}
  \renewcommand{\arraystretch}{1.2}
  \begin{tabular}{>{\centering\arraybackslash}m{2cm}>{\centering\arraybackslash}m{2cm}>{\centering\arraybackslash}m{3cm}>{\centering\arraybackslash}m{3cm}>{\centering\arraybackslash}m{2cm}>{\centering\arraybackslash}m{2cm}}
    \hline
   \textit{Number of Requests} & \textit{Create DID Documents} & \textit{Create Insurance Contract} & \textit{Update Insurance Contract Signature} & \textit{Create Claim} & \textit{Update Claim} \\ \hline\hline
    50  & 0     & 0     & 0     & 0  & 0  \\ \hline
    100 & 0     & 0     & 0     & 0  & 0  \\ \hline
    250 & 0     & 0     & 0     & 0  & 0  \\ \hline
    500 & 0     & 0     & 0     & 0  & 0  \\ \hline
    1000 & 0    & 0     & 0     & 0  & 0  \\ \hline
    1500 & 0    & 0     & 0     & 0  & 0  \\ \hline
    2000 & 0    & 0     & 0     & 0  & 0  \\ \hline
    2500 & 0    & 12  & 12  & 16 & 20 \\ \hline
    3000 & 0    & 16  & 16  & 20  & 30 \\ \hline
  \end{tabular}
\label{tab:ErrorRateTable}
\end{table*}


Overall, the results show that higher concurrency leads to higher latency for all evaluated operations. Throughput increases with load in most cases, but its growth slows or fluctuates at higher concurrency levels, suggesting that the system approaches saturation. Dropped connections appear from 2500 concurrent requests onward for all operations except Create DID Document.

The increase in latency is mainly explained by finite host resources being shared among a growing number of concurrent requests. Although throughput does not collapse within the tested range, the error rates in Table \ref{tab:ErrorRateTable} show that the system becomes less reliable under the highest loads. These results should therefore be interpreted as prototype-level performance measurements on a resource-limited local deployment, rather than as production-scale benchmarks.

\section{Conclusion}

This paper presents a blockchain-based solution for the insurance market to streamline the registration of insurance contracts and the handling of claims with enhanced security through blockchain technology and digital signatures. The solution combines a permissioned blockchain for shared record-keeping with IPFS for decentralized file storage, while employing satellite imagery for property damage assessment.

The system is built on a private, permissioned Hyperledger Fabric blockchain, enabling identity registration for different entities and using IPFS for content-addressed storage. The blockchain stores three types of assets: Decentralized Identifier (DID) Documents, Insurance Contracts, and Damage Claims. Four participants were defined: Client, Insurance Company, Auditor, and Higher Authority, each with specific permissions. Authentication is achieved through DID documents and Fabric's Membership Service Provider (MSP).

The desktop app provides access to all features, with authentication handled in each transaction using uploaded credentials. The evaluation of the system included measurements of latency, throughput, and error on a resource-limited local machine. The results show increasing throughput up to high concurrency levels, at the cost of higher latency and dropped connections beyond 2500 concurrent requests.

In summary, this solution improves transparency and efficiency in the insurance process, ensuring that all parties have secure, authenticated access to relevant information, while leveraging blockchain for an immutable record of transactions.

\section*{Acknowledgements} 
This work was financially supported by Project Blockchain.PT – Decentralize Portugal with Blockchain Agenda (Project no 51), WP 6: Digital Assets Management, Call no 02/C05-i01.01/2022, funded by the Portuguese Recovery and Resilience Program (PRR), The Portuguese Republic, and The European Union (EU) under the framework of the Next Generation EU Program. This work was also supported by national funds through Fundação para a Ciência e a Tecnologia, I.P. (FCT) under projects UID/50021/2025 (DOI: https://doi.org/10.54499/UID/50021/2025).

\bibliographystyle{IEEEtran}
\bibliography{bib.bib}

@String{Computing = "Computing" }

@String{Computer = "{IEEE} Computer" }

@String{Springer = "Springer-Verlag" }

@article{dinh2018untangling,
  title={Untangling blockchain: A data processing view of blockchain systems},
  author={Dinh, Tien Tuan Anh and Liu, Rui and Zhang, Meihui and Chen, Gang and Ooi, Beng Chin and Wang, Ji},
  journal={IEEE Transactions on Knowledge and Data Engineering},
  volume={30},
  number={7},
  pages={1366--1385},
  year={2018}
}

@article{nakamoto2009bitcoin,
  author = {Nakamoto, Satoshi},
  title = {Bitcoin: A Peer-to-Peer Electronic Cash System},
  year = 2008
}

@unpublished{buterin2014ethereum,
	author = {Vitalik Buterin and {Ethereum team}},
	note = {White paper},
	title = {Ethereum: A Next-Generation Smart Contract and Decentralized Application Platform},
	year = {{2014-17}}}

@article{reed2020decentralized,
  title={Decentralized identifiers ({DIDs}) v1. 0},
  author={Reed, Drummond and Sporny, Manu and Longley, Dave and Allen, Christopher and Grant, Ryan and Sabadello, Markus and Holt, Jonathan},
  journal={Draft Community Group Report},
  year={2020}
}

@Inbook{Dhillon2017,
author="Dhillon, Vikram
and Metcalf, David
and Hooper, Max",
title="The Hyperledger Project",
bookTitle="Blockchain Enabled Applications: Understand the Blockchain Ecosystem and How to Make it Work for You",
year="2017",
publisher="Apress",
address="Berkeley, CA",
pages="139--149"
}

@techreport{w3c-vcdata-model,
  author = {Manu Sporny and Dave Longley and David Chadwick},
  title = {Verifiable Credentials Data Model 1.0},
  institution = {W3C},
  year = {2019},
  pages = {1--115},
  url = {https://w3c.github.io/vcdata-model/https://www.w3.org/TR/vc-data-model/},
}

@article{shuaib2020blockchain,
  title={Blockchain-based framework for secure and reliable land registry system},
  author={Shuaib, Mohammed and Daud, Salwani Mohd and Alam, Shadab and Khan, Wazir Zada},
  journal={TELKOMNIKA (Telecommunication Computing Electronics and Control)},
  volume={18},
  number={5},
  pages={2560--2571},
  year={2020}
}

@online{ICA-IT:BlockchainLandregistry,
  title = {Blockchain and Land Registry},
  author = {{International Council for Accreditation of {IT} Managers}},
  year = {2016},
  url = {https://ica-it.org/pdf/Blockchain\_Landregistry\_Report.pdf}
}

@online{BlockchainGovIndia:LandRegistration,
  title = {Land Registration Using Blockchain},
  author = {{Government of India, Ministry of Electronics and Information Technology}},
  year = {2020}, 
  url = {https://blockchain.gov.in/Home/CaseStudy?CaseStudy=LandRegistration}
}

@online{BorgenProject:BlockchainLandRegistry,
  title = {Blockchain-Based Land Registry},
  author = {{Borgen Project}},
  year = {2021},
  url = {https://borgenproject.org/blockchain-based-land-registry/}
}

@online{NikkeiAsia:JapanPropertyRecords,
  title = {Japan to Tidy Up Scattered Property Records},
  author = {Nikkei Asia},
  year = {2017},
  url = {https://asia.nikkei.com/Markets/Property/Japan-to-tidy-up-scattered-property-records}
}

@article{lemieux2018title,
  title={Title and code: Real Estate Transaction Recording in the Blockchain in {Brazil (RCPLAC-01)}},
  author={Lemieux, Victoria and Lacombe, Claudia and Flores, Daniel},
  journal={Case Study},
  volume={1},
  year={2018}
}

@article{khan2020blockchain,
  title={Blockchain based land registry system using Ethereum Blockchain},
  author={Khan, Rijwan and Ansari, Shadab and Jain, Sneha and Sachdeva, Saksham},
  journal={Journal of Xi'an University of Architecture \& Technology},
  volume={12},
  pages={3640--3648},
  year={2020}
}

@inproceedings{sahai2020smart,
  title={Smart contract definition for land registry in blockchain},
  author={Sahai, Archana and Pandey, Rajiv},
  booktitle={IEEE 9th International Conference on Communication Systems and Network Technologies},
  pages={230--235},
  year={2020}
}

@article{benet2014ipfs,
  title={{IPFS}: Content addressed, versioned, {P2P} file system},
  author={Benet, Juan},
  journal={arXiv preprint arXiv:1407.3561},
  year={2014}
}

@article{peck2017blockchains,
	author = {Peck, Morgen E},
	journal = {IEEE Spectrum},
	number = {10},
	pages = {26--35},
	title = {Blockchains: How they work and why they'll change the world},
	volume = {54},
	year = {2017}}

@inproceedings{RAFTongaro2014search,
  title={In search of an understandable consensus algorithm},
  author={Ongaro, Diego and Ousterhout, John},
  booktitle={2014 USENIX Annual Technical Conference},
  pages={305--319},
  year={2014}
}

@inproceedings{cohen2003incentivesBITTORRENT,
  title={Incentives build robustness in BitTorrent},
  author={Cohen, Bram},
  booktitle={Workshop on Economics of Peer-to-Peer systems},
  volume={6},
  pages={68--72},
  year={2003}
}

@misc{swarm_whitepaper,
  title        = {Swarm: The Decentralized Storage and Communication System for a Sovereign Digital Society},
  author       = {{Swarm Foundation}},
  howpublished = {\url{https://www.ethswarm.org/swarm-whitepaper.pdf}},
  year         = 2020,
  note         = {Accessed: 2024-08-16}
}

@misc{hyperledger_fabric_2024,
  title        = {Hyperledger {Fabric} Documentation},
  author = {{Hyperledger Fabric team}},
  year         = {2024},
  url          = {https://hyperledger-fabric.readthedocs.io/en/release-2.5/},
  note         = {Accessed: 2024-10-08}
}

@manual{verifiable_credential_2023,
	author = {Manu Sporny and Ted Thibodeau Jr and Ivan Herman and Michael B. Jones and Gabe Cohen},
	month = {oct},
	organization = {W3C},
	title = {Verifiable Credentials Data Model v2.0 ({W3C} Candidate Recommendation Draft)},
	year = {2024}}

@article{hardman2022sovereignidentity,
  title={The Path to Self-Sovereign Identity},
  author={Christopher Allen},
  year={2016},
  journal={Life With Alacrity},
  url={https://www.lifewithalacrity.com/article/the-path-to-self-soverereign-identity/}
}

@article{CAPTHEOREM,
  title={Perspectives on the {CAP} Theorem},
  author={Gilbert, Seth and Lynch, Nancy},
  journal={Computer},
  volume={45},
  number={2},
  pages={30--36},
  year={2012},
  publisher={IEEE}
}

@inproceedings{henriques2025decentralised,
  title={Decentralised Land Registration and Transaction with Blockchain and Self-Sovereign Identity},
  author={Henriques, Pedro C and Correia, Miguel},
  booktitle={2025 44th International Symposium on Reliable Distributed Systems (SRDS), DLT4SEC 2025 - International Workshop on DLT for Cybersecurity and Vice Versa},
  pages={404--409},

  year={2025},
  organization={IEEE}
}

@article{dominguez2024blockchainInsurance,
  author  = {Dominguez Anguiano, Teresa and Parte, Laura},
  title   = {The State of Art, Opportunities and Challenges of Blockchain in the Insurance Industry: A Systematic Literature Review},
  journal = {Management Review Quarterly},
  volume  = {74},
  number  = {2},
  pages   = {1097--1118},
  year    = {2024},
  publisher={Springer}
}

@article{farao2024inchain,
  author  = {Farao, Aristeidis and Paparis, Georgios and Panda, Sakshyam and Panaousis, Emmanouil and Zarras, Apostolis and Xenakis, Christos},
  title   = {{INCHAIN}: A Cyber Insurance Architecture with Smart Contracts and Self-Sovereign Identity on Top of Blockchain},
  journal = {International Journal of Information Security},
  volume  = {23},
  pages   = {347--371},
  year    = {2024},
  publisher={Springer}
}

@article{chen2023blockchainIPFSCarInsurance,
  author  = {Chen, Chin-Ling and Zheng, Ying-Ming and Huang, Der-Chen and Liu, Ling-Chun and Chen, Hsing-Chung},
  title   = {A Blockchain and {IPFS}-Based Anticounterfeit Traceable Functionality of Car Insurance Claims System},
  journal = {Sensors},
  volume  = {23},
  number  = {23},
  pages   = {9577},
  year    = {2023}
}

@article{steinmann2023naturalHazardParametric,
  author  = {Steinmann, Carmen B. and Guillod, Beno{\^i}t P. and Fairless, Christopher and Bresch, David N.},
  title   = {A Generalized Framework for Designing Open-Source Natural Hazard Parametric Insurance},
  journal = {Environment Systems and Decisions},
  volume  = {43},
  number={4},
  pages={555--568},
  year={2023},
  publisher={Springer}
}

@article{rabehaja2024parametricInsurance,
  author  = {Rabehaja, Tahiry and Pal, Shantanu and Hill, Ambrose and Hitchens, Michael},
  title   = {A Blockchain-Based Approach for Parametric Insurance Under Multiple Sources of Truth},
  journal = {IEEE Transactions on Services Computing},
  volume  = {17},
  number  = {3},
  pages   = {718--732},
  year    = {2024},
  doi     = {10.1109/TSC.2023.3296808},
  url     = {https://doi.org/10.1109/TSC.2023.3296808}
}

@article{hao2025privacy,
  title={Privacy-preserving blockchain-enabled parametric insurance via remote sensing and {IoT}},
  author={Hao, Mingyu and Qian, Keyang and Chau, Sid Chi-Kin},
  journal={IEEE Transactions on Services Computing},
  volume={18},
  number={5},
  pages={3093--3105},
  year={2025},
  publisher={IEEE}
}

@article{liu2024remoteSensingVerification,
  author  = {Liu, Yujie and Chang, Yuanfei},
  title   = {Blockchain-Based Method for Spatial Retrieval and Verification of Remote Sensing Images},
  journal = {Sensors},
  volume  = {24},
  number  = {7},
  year    = {2024}
}

\end{document}